# Gaming the System: Video Games as a Theoretical Framework for Instructional Design

*Ian D. Beatty, University of North Carolina at Greensboro*

**Abstract:** In order to facilitate analyzing video games as learning systems and instructional designs as games, we present a theoretical framework that integrates ideas from a broad range of literature. The framework describes games in terms of four layers, all sharing similar structural elements and dynamics: a *micro-level game* focused on immediate problem-solving and skill development, a *macro-level game* focused on the experience of the game world and story and identity development, and two *meta-level games* focused on building or modifying the game and on social interactions around it. Each layer casts gameplay as a co-construction of the game and the player, and contains three dynamical feedback loops: an exploratory learning loop, an intrinsic motivation loop, and an identity loop.

## Introduction

Video games are of increasing interest to scholars in the Learning Sciences. They are emerging as a major literacy in our society, especially in the lives of our students (Koster, 2005 p.viii, forward by Will Wright). They are powerful, with affordances far beyond those of other media for engaging, for persuading, and for educating (Bogost, 2007; McGonigal, 2011). In fact, they already teach; Gee (2007b) has argued that good video games are in fact carefully engineered learning machines:

> Game companies face an interesting problem, a problem that schools face, as well: how to get someone to learn something that is challenging and requires persistence… If people can't learn to play a company's games, the company goes broke. So game designers have no choice, they have to make games that are very good at getting themselves learned. (p. 2)

Players of such games can learn scientific thinking skills (Steinkuehler & Duncan, 2008), develop capacities for teamwork and social coordination (McGonigal, 2011), and gain insight into professions, social contexts, and moral dilemmas otherwise inaccessible to them (Gee, 2007a). Through a mechanism that Gee (2005) calls *distributed authentic professionalism*, players can experience the practice of a "profession," *becoming* a new kind of person while learning some of the skills, values, and ways of seeing of that profession.

For those of us interested in improving school-based instruction, recognizing the learning potential of video games leads to three possible courses of action. One is developing games that teach our subject content. A second is "gamifying" standard instruction, dressing it up in the superficial trappings of games. A third course, which I advocate, is learning from the design of effective video games, and using what we learn to reconceptualize teaching and design novel approaches to classroom instruction. To pursue such a course, we need theoretical tools to support thinking about, communicating about, and researching games as learning environments and learning environments as games.

In response to this need, I am developing a comprehensive theoretical framework that draws on and integrates insights from disparate literature bases: practical advice on effective game design by successful designers, scholarly analyses of the learning features and potential of games by learning scientists, and empirical reports of game-based and game-inspired teaching experiments by innovative instructors. The resulting framework can be used as a lens to support theoretical analysis of games and learning, to support practical instructional design, and to support empirical studies of learning environments. This paper presents a brief and high-level summary of the framework.

## Background

As leading game designers have observed (Koster, 2005; Schell, 2008), the essence of a "game" is puzzle-solving or problem-solving, in which competence is developed through an exploratory, trial-and-error learning dynamic. Players must choose among possible actions within some simplified and iconified game world, experience the consequences of their choices, learn from those consequences and the feedback provided by the game, and iterate. Most games take players through a sequence of challenges of increasing difficulty. The root of games' remarkable motivating power is the thrill of challenge, learning, and mastery. As Koster puts it, "Fun from games arises out of mastery. It arises out of comprehension. It is the act of solving puzzles that makes games fun. In other words, with games, learning is the drug" (2005, p. 40).

McGonigal, in a sweeping assessment of the factors that make games so popular and the ways in which they serve or can be made to serve productive ends, identifies four defining traits of a "game" (2011, p. 21). The first is a *goal*, which focuses players' attention, continually orients their participation, and provides a sense of purpose. The second is the *rules*, which impose artificial constraints to "push players to explore previously uncharted possibility spaces" and "unleash creativity and foster strategic thinking." The third is the *feedback system*, which "tells players how close they are to achieving the goal… Real-time feedback serves as a promise to the player that the goal is definitely achievable, and it provides motivation to keep playing." The fourth is *voluntary participation*, which "ensures that intentionally stressful and challenging work is experienced as safe and pleasurable activity."

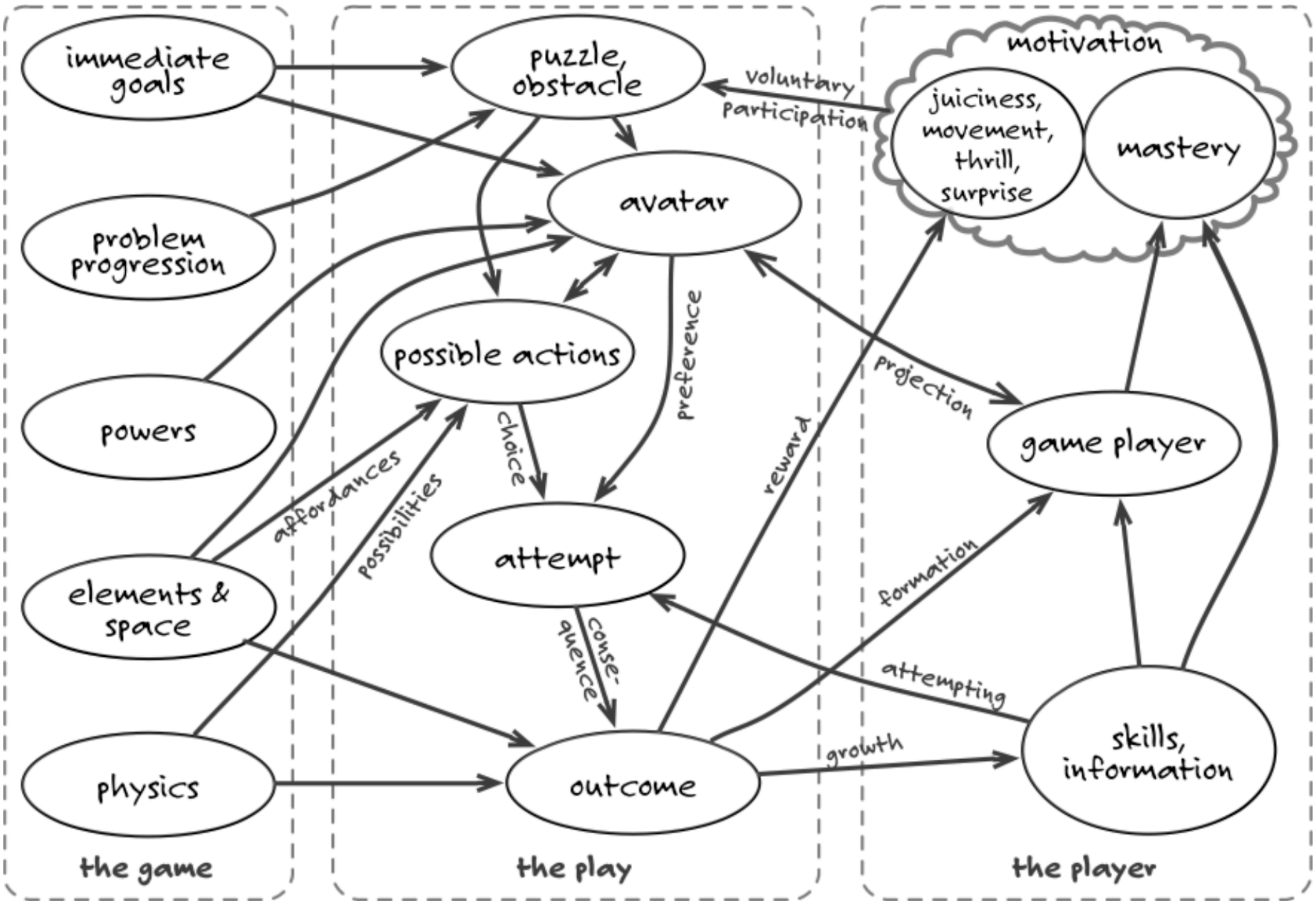

Figure 1. The *micro-level game* layer, focusing on problem-solving and on learning facts and skills.

Gee, a sociocultural linguist turned video game scholar, claims that:

> [T]he designers of many good games have hit on profoundly good methods of getting people to learn and to enjoy learning. They have had to, since games that were bad at getting themselves learned didn't get played and the companies that made them lost money. Furthermore, it turns out that these learning methods are similar in many respects to cutting-edge principles being discovered in research on human learning. (2007b, p. 29)

Based on careful analysis of many different video games, he goes on to articulate thirteen general "good principles of learning built into good computer and video games" (p. 30). These include, among others, *co-design*: "Good learning requires that learners feel like active agents (producers) not just passive recipients (consumers)"; *identity*: "Deep learning requires an extended commitment and such a commitment is powerfully recruited when people take on a new identity they value and in which they become heavily invested…"; *cycles of expertise*: "Expertise is formed in any area by repeated cycles of learners practicing skills until they are nearly automatic, then having those skills fail in ways that cause the learners to have to think again and learn anew…"; *information on demand and just in time*: "Human beings… use verbal information best when it is given 'just in time' (when they can put it to use) and 'on demand' (when they feel they need it)"; *skills as strategies*: "People learn and practice skills best when they see a set of related skills as a strategy to accomplish goals they want to accomplish"; and *meaning as action image*: "Humans… think [by way of] experiences they have had and imaginative reconstructions of experience… Words and concepts have their deepest meanings when they are clearly tied to perception and action in the real world."

Gee (2005, 2007a) also argues that video games allow and encourage players to explore and develop new identities, and that such identity development is a powerful component of learning. He claims that learning physics, anthropology, urban planning, medicine, or military strategy means (or should mean) learning to *be* a physicist, anthropologist, urban planner, doctor, or military strategist, and that learning to be a professional means learning the profession's "special ways of acting and interacting in ways that produce and use the domain's knowledge… special ways of seeing, valuing, and being in the world" (2005, p. 1). Through an approach he calls *distributed authentic professionalism*, video games are particularly effective at helping players assume a new identity, performing virtual actions and having virtual experiences that they are not yet capable of in the real world. "By the end of the game, the player has experienced a 'career' and has a story to tell about how his or her professional expertise grew and was put to tactical and strategic uses" (2005, p. 4).

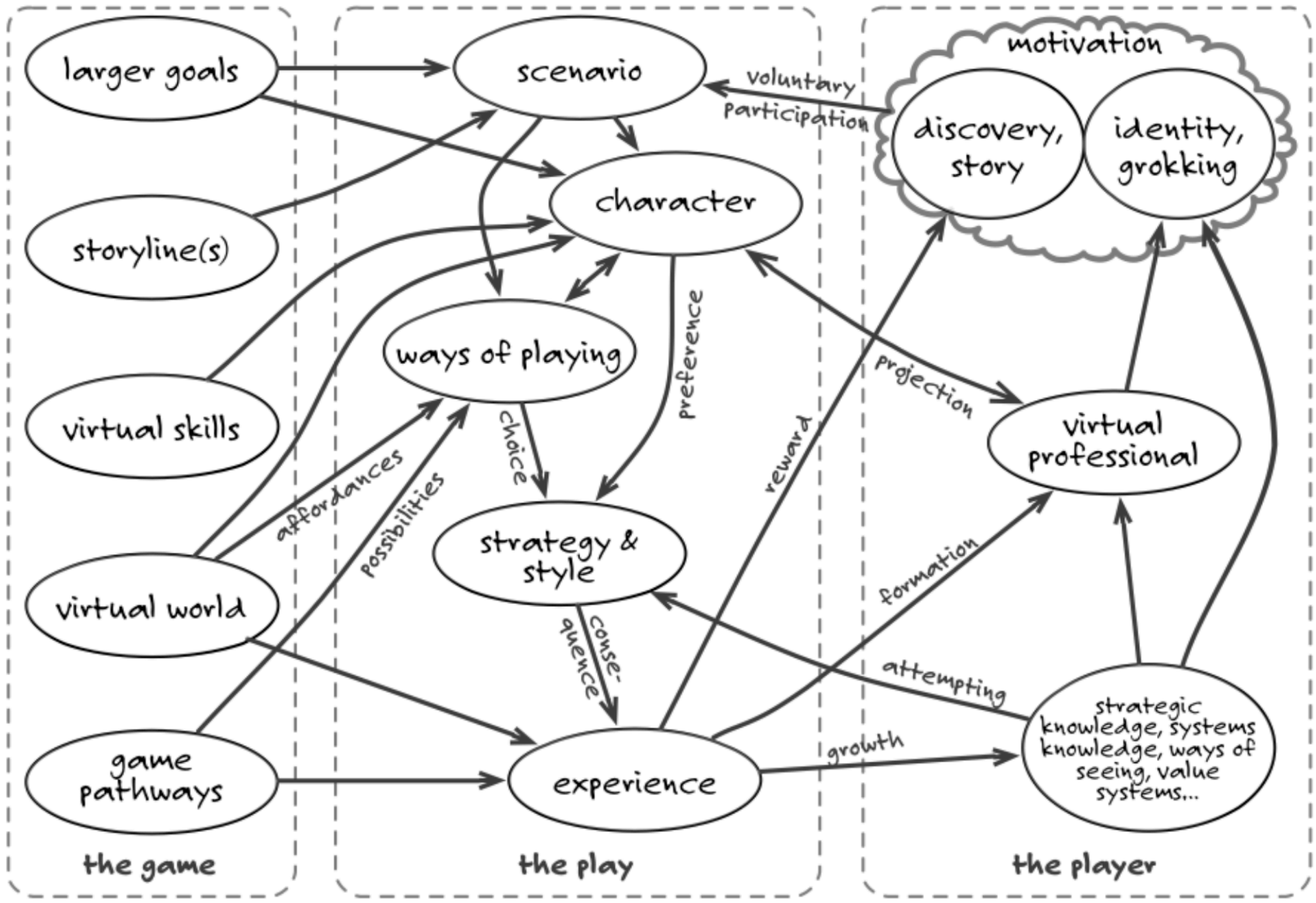

Figure 2. The *macro-level game* layer, focusing on the experience of the world and storyline, and on learning "professional" attitudes and identity.

Squire (2006) observes that games are multi-layer phenomena. For complex modern video games, the dynamic of immediate problems, choices, actions, consequences, and iteration is set with in the larger context of a *designed experience*. In this context, players experience a game world and storyline that communicate an ideology and develop players' understandings through "cycles of performance" and "a grammar of doing and being" (p. 19). That is, the dynamic of skill learning gives rise to a dynamic of in-game identity development. Squire (2006) and Steinkuehler (2004, 2006) identify another important facet of game-play: the "affinity space" (Gee, 2007) of inter-player social interactions that occurs within and around many modern games, giving rise to a variety of social structures, shaping game-play strategies and identities, and connecting players' game identities to their real-life ones: a meta-game dynamic of identity development. Additionally, as Gee (2007) notes, many games allow players to become designers and builders by providing tools for them to modify or extend portions of the game-world. This enables another meta-game of designing and contribution.

## The Framework

### Layers and Elements

To synthesize and detail these aspects of game-based learning, as well as many others consonant with them, I propose the following multi-layer framework. It contains four layers, each describing a different dimension or scale of the game-playing experience. All layers follow the same general template, revealing that the same fundamental dynamical patterns and learning principles are at work in each of these levels.

The framework's *micro-level game* layer (Figure 1) describes a game at the level of the specific challenges or obstacles that a player must solve, the nature of the "avatar" she controls, the actions possible to that avatar, the specific attempts she makes, and the resulting outcomes. Trial-and error exploratory learning leads to a growth in the player's skills and knowledge of game information. Motivation is maintained through both the attractions of the game itself and, more importantly, the pleasure of mastering challenges.

In order to describe the "designed experience" (Squire, 2006) that emerges from this micro-level game, the framework contains a second, *macro-level game* layer (Figure 2). This layer describes the fictional setting and scenario of the game, the outlook and motivations of the character representing the player, the various ways of playing the game available, the gameplay strategy and style adopted by the player, and the resulting experience that the game and player co-construct. Exploration of the game world and accumulating experiences lead to a growth in the player's strategic and systems knowledge of the game world's dynamics, and shape her value systems and ways of seeing as she develops an identity as a virtual "professional" (i.e., functional expert in something). Motivation arises from enjoyment of the unfolding story and environment, as well as from the satisfactions of "grokking" (deeply and holistically understanding) something and of exploring and growing an identity.

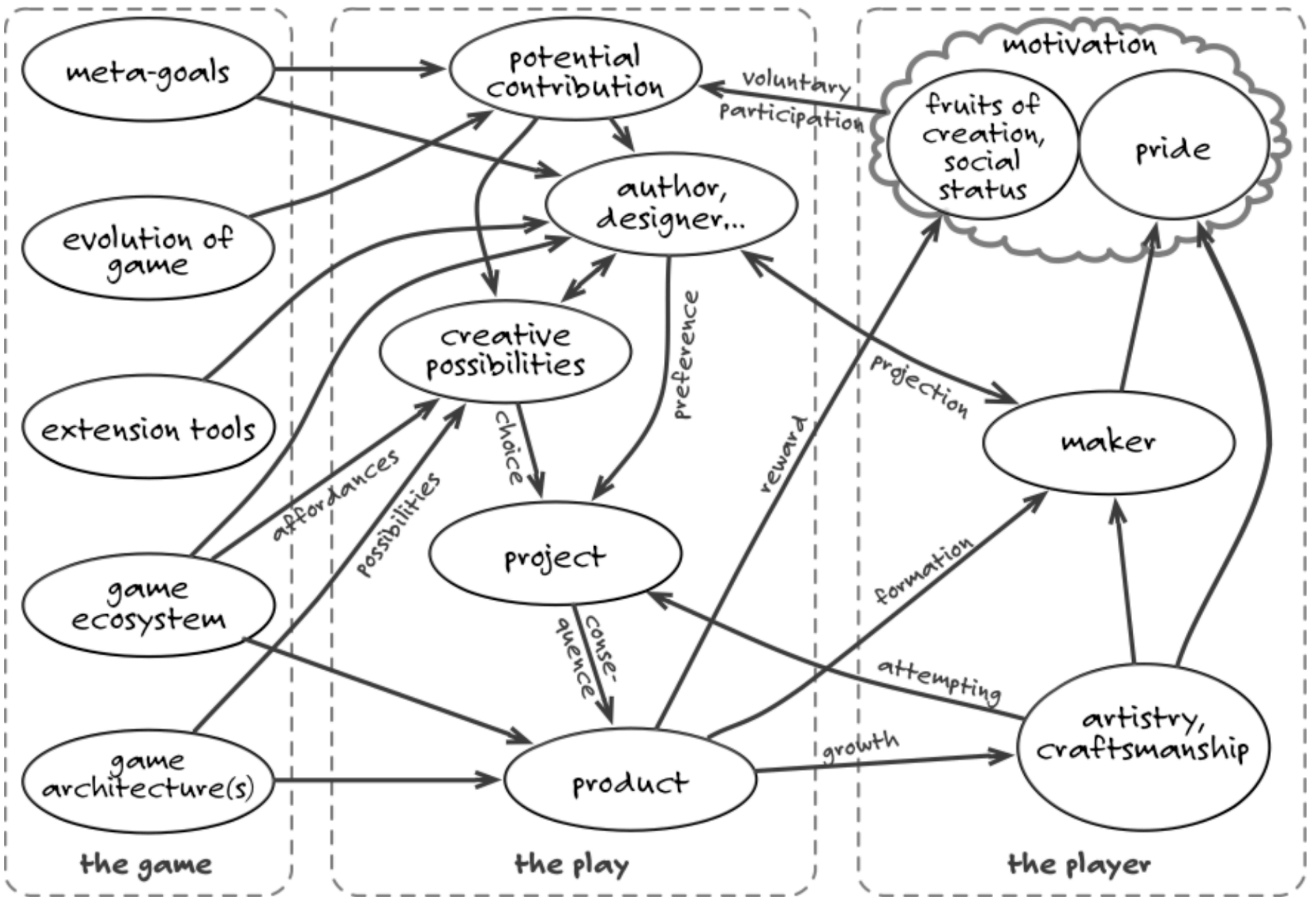

Figure 3. The *builder meta-level game* layer, focusing on contribution to the game, and on learning design skills.

In order to capture the meta-game experiences that take place between the players of a game and often constitute an essential component of game-play, the framework includes two additional layers. The *builder meta-game* layer (Figure 3) describes the dynamics through which players can use game-provided or third-party tools to extend, modify, augment, or document the game. It describes the potential contributions the player can make, the creative possibilities open, the project(s) that the player undertakes, and the product(s) that result. Engaging in such authoring or design work develops the player's artistry or craftsmanship, developing her identity as a maker rather than just a consumer. Motivation arises from the desire to enjoy the fruits of creation—that is, the specific products created and the social status that may result—as well as from pride in accomplishment.

The framework's *social meta-game* layer (Figure 4) describes social interactions within the "affinity space" of individuals brought together by a common interest in the game. It describes the collective action that groups may engage in (perhaps goal-directed, perhaps merely to socialize), the roles that individuals may assume within the society and the social possibilities open to them, the social moves (actions or gambits) the player makes, and the resulting interactions that she experiences. Through such social interaction, the player develops social connections and roles, and potentially becomes a valued member of the community. Motivation can arise from the pleasures of social interaction as well as from feelings of connection and significance.

All four of these levels follow the same *generic layer template* (Figure 5). Faced with a problem or challenge that the player has voluntarily chosen (by opting into the game or meta-game activity), the player's virtual identity within the game or affinity space chooses from among available options to attempt a solution. The rules of the environment lead to a result and feedback to the player. By exploring possible choices and learning from feedback, the player grows relevant capacities and develops aspects of her identity. External motivation arises from the direct experience and results of participating in the game or meta-game, while internal motivation arises from the experience of challenge and mastery and of identity growth. The fact that each layer shares a common structure is perhaps not surprising: this common structure, I claim, illuminates the fundamental dynamics of game-play and the learning principles they leverage that give games their power to entertain and to educate. These dynamics permeate a game at all its levels, including its basic mechanics, its higher-level narrative, and its emergent meta-game activities. To put it another way, the aspects of the game that are emergent from the base game share and echo the base game's characteristics.

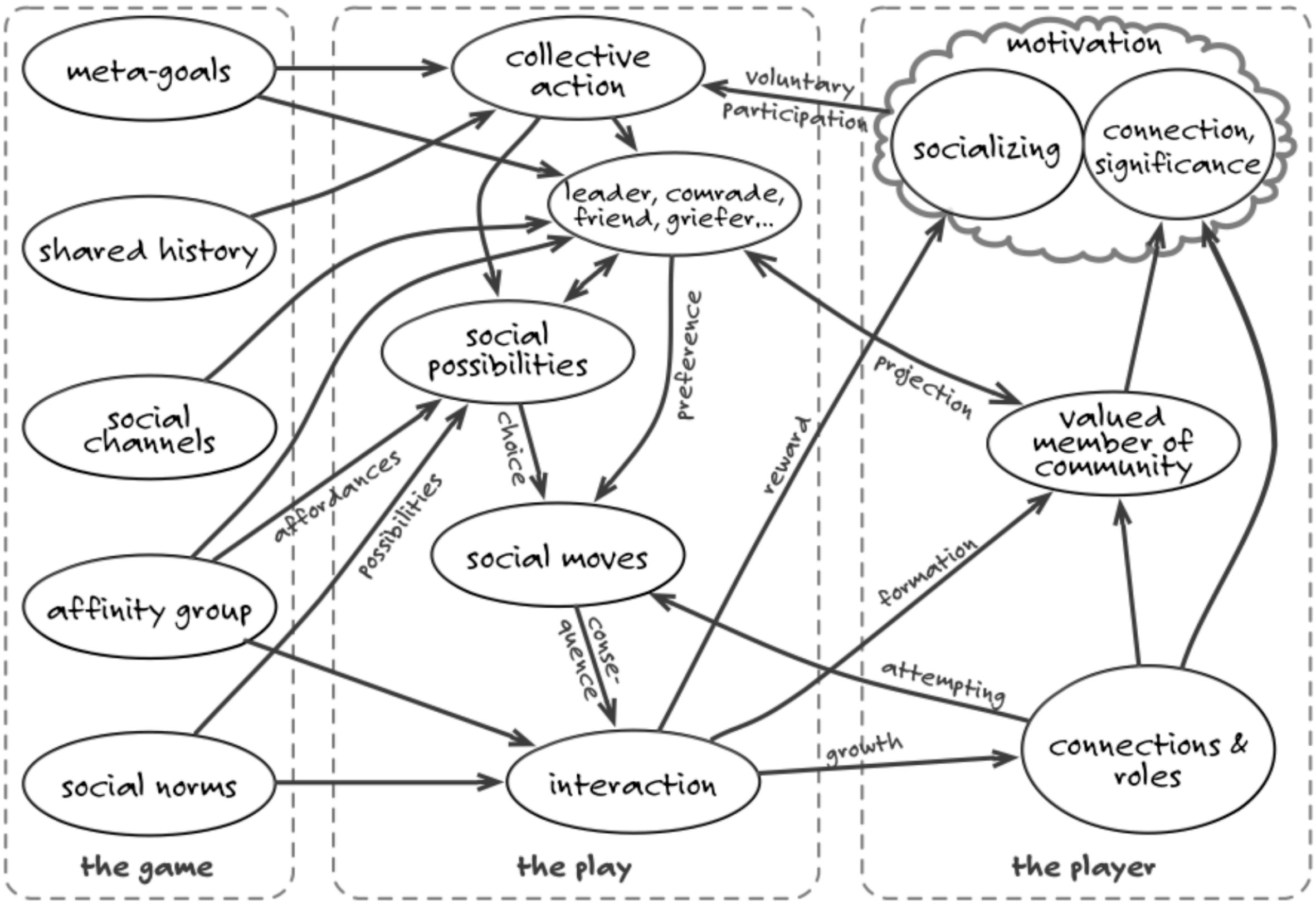

Figure 4. The *social meta-level game* layer, focusing on social experiences within the affinity group, and on forming a meta-game social identity.

## Implicit Features

Some of these "fundamental dynamics" are not captured in specific named elements (i.e., layers, ovals, or arrows), but rather in their organization and patterns of interconnection. For example, Gee's *co-design* principle—that "Good learning requires that learners feel like active agents (producers) not just passive recipients (consumers)"—is represented in the generic framework template and each layer by the fact that "the play" is a co-construction of "the game" and "the player." (As Schell, 2008, points out, this is perhaps the primary reason that video games can provide a much more powerful experience than movies or books, and also why they are so much more difficult to craft.) This principle is also implicit in the fact that the entire system's dynamics are driven by player choice: to accept challenges and problems, to choose actions in the absence of expertise, and to iterate and explore until succeeding. *Player agency* is deeply woven into the fabric of games, and therefore of the framework.

Other fundamental dynamics are represented by feedback loops among subsets of a layer's elements. Figure 6 shows two such loops: the *exploratory learning loop* and the *intrinsic motivation loop*. The *exploratory learning loop* captures the fundamental dynamic common to almost all games: learning by doing—or, rather, learning by trying, failing, reassessing, and trying again repeatedly until succeeding. This requires that failure be safe and informative, and that possible avenues of action are evident to or discoverable by the player.

The *intrinsic motivation loop* captures the fact that while good games may provide players with some spectacle or other pleasures that generate extrinsic motivation, their primary motivating power comes from the deeply intrinsic satisfaction of undertaking and eventually succeeding at the challenges they offer. For most players, learning is its own reward, at least in the context of a good game (whether that be a video game or an athletic sport).

Figure 7 shows another such feedback loop, the *identity loop*. Gee (2007a) analyzes the complex identity dynamics of a video game by asserting three different kinds of identity that interact. The first is the player's *real-world identity*, possessing the real-life capacities, values, and perceptions that the player brings to the game. The second is the player's *virtual identity*, her fictional persona in the game-world with its in-game representation, capacities, goals, and values. The third is a *projective identity*, in which the player projects her own values and desires onto the virtual character, and also sees the virtual character "as my own project in the making, a creature whom I imbue with a certain trajectory through time defined by my aspirations for what I want that character to be and become" (p. 50). Through the projection of the player's real-world identity into the game world's virtual identity, the game becomes a venue for exploring and potentially growing her real-world identity. This can become very powerful in the meta-game layers, where resonance with the identity dynamics of other players is possible.

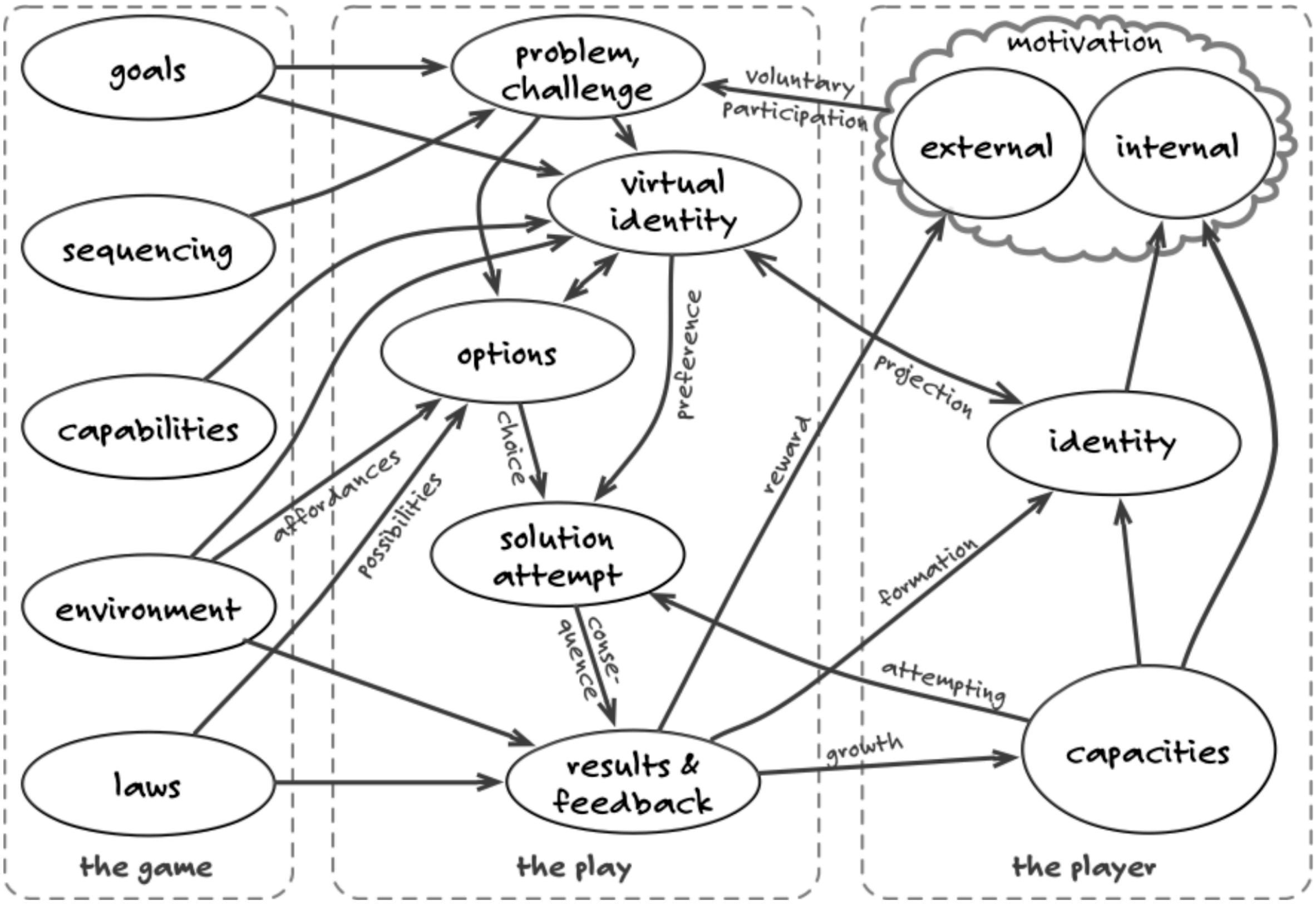

Figure 5. The generic template for a framework layer.

## Discussion

### Contrasts

If we take seriously the notion that video games are powerful learning systems—whether or not the "content" most extant games teach is particularly valuable—and that, as Gee says, game designers "have hit on profoundly good methods of getting people to learn and to enjoy learning" (2007b, p. 29), we should find this framework quite provocative. By representing games as learning systems, it draws our attention to many ways in which game-based learning is profoundly different from most school-based learning.

One difference is that good games feed a player's sense of agency in many ways: they often let players choose from among challenges to undertake and select how they will approach them, on both tactical and strategic levels. They often let players design or customize their virtual character, develop a personalized game persona, and select from among different styles of play. The unfolding game itself is co-constructed with the player, so that the unfolding storyline of the macro-game layer seems like a natural consequence of the player's choices. Often, the player's actions are of epic significance within the game world. In the meta-game layers, a player can reshape or extend the game, and can parlay game activity and skill into social status and connections. The fact that the framework's initial connection from "the player" to "the play" begins with "voluntary participation" is of tremendous significance: The player can always choose to walk away from the game, temporarily or permanently. In some sense, that gives her absolute power over it. Meanwhile, most school-based learning requires students to learn the content that educators specify, in the order specified, on a designated pace, using the specific texts and activities provided. They must demonstrate learning in specific ways. Schools rely heavily on external motivators such as grade threat. Rather than encouraging students to explore their own identities and find creative ways to invest those in the learning, educators all to often steer students to adopt an "ideal student" role.

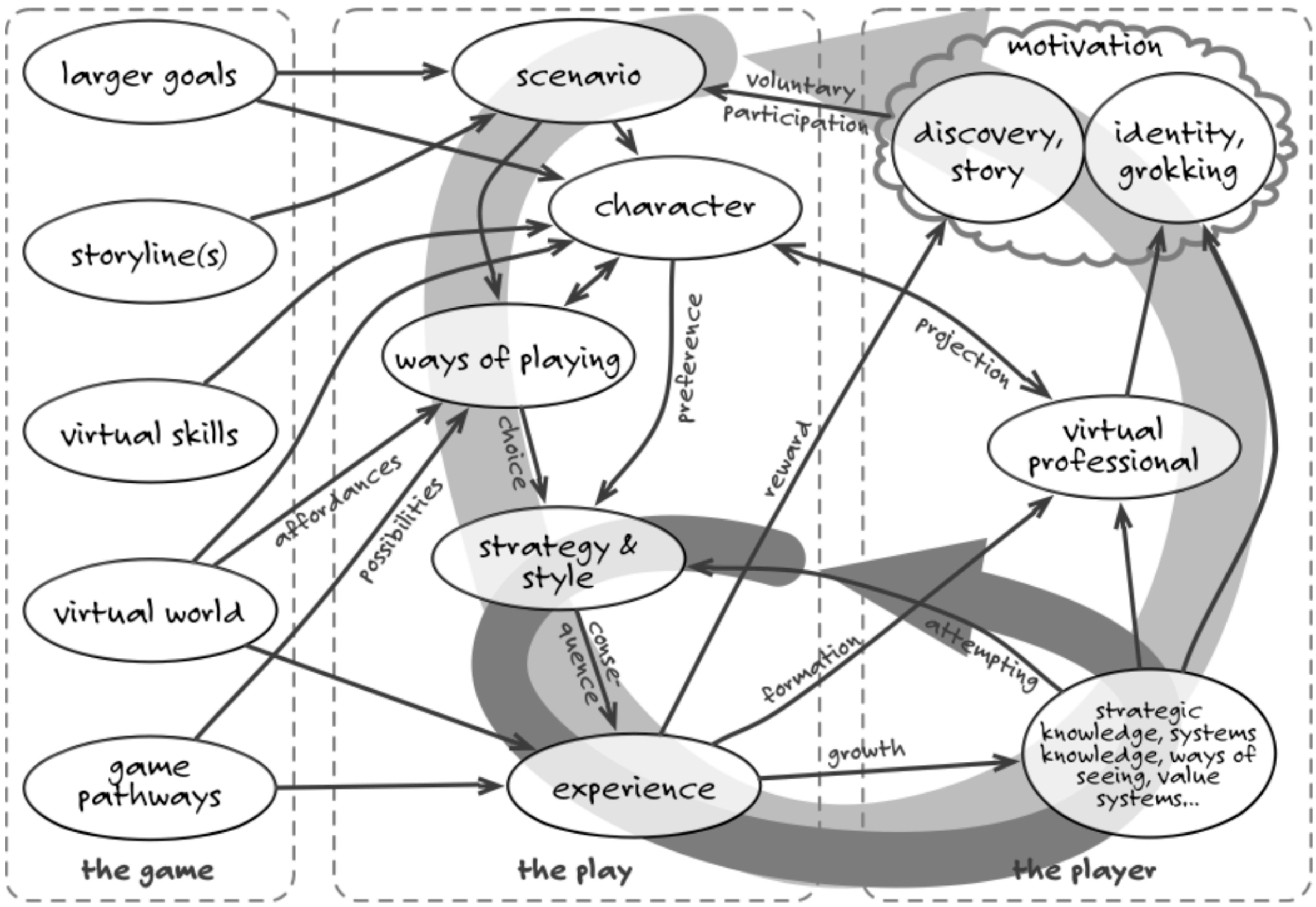

Figure 6. The *exploratory learning loop* (dark gray, smaller) and the *intrinsic motivation loop* (light gray, larger) shown on the generic framework layer.

A second difference is that in games, the fundamental learning dynamic is one of trial-and-error exploration. Multiple courses of action are available and evident. Failure is safe and often entertaining, generally meaning no more than "you haven't succeeded—yet." Feedback is immediate and informative. "Content" information is provided just-in-time and on demand, when players can immediately apply it or request it. Tasks are sequenced so that players can develop a sense of mastery with one set of skills before being pushed by greater challenges to develop new ones. In contrast, typical school courses permit a single attempt at any given challenge, and failures permanently weigh down a student's course grade. (Some novel grading approaches, such as "standards-based grading," seek to avoid this; cf. Beatty, 2013 and references therein.) Feedback is more evaluative than constructive, and opportunities to apply that feedback are often scarce.

A third difference is that good games deliberately recruit players' investment in a novel identity, and take advantage of the medium to provide players with the experience of being a different kind or profession of person than is possible in real life. Most school-based instruction, on the other hand, is more focused on teaching skills and content knowledge than on nurturing the values and ways of seeing the world that give meaning to those skills and that knowledge.

## Possibilities

In addition to using the framework to critique extant instructional designs, we can use it generatively to suggest alternative ones. For example, the process of developing this framework brought to my attention the importance to learning of natural, informative, rapid feedback: feedback that arises "naturally" as a consequence of the learner's actions (rather than being imposed by an authority), directs the learner's attention to the specific choices or abilities that need improvement and reveals how they were inadequate, and occurs quickly enough to support an efficient trial-and-error learning process. As a result, I modified the *Introduction to Computational Physics* course I teach in order to provide more such feedback. One change was to craft programming assignments whose success or failure would be more immediately obvious to the students, rather than requiring my evaluation. A second was to provide students with automated "test code" that would quickly put their programs through a battery of tests and display diagnostic results.

Similarly, the importance of "perceptible options" in the framework—so that players always see something potentially productive to try—inspired me to compile an explicit list of programming "tools" and tactics for each unit, phrased in ways that help students think of them as building blocks of strategies to solve the unit's programming challenges.

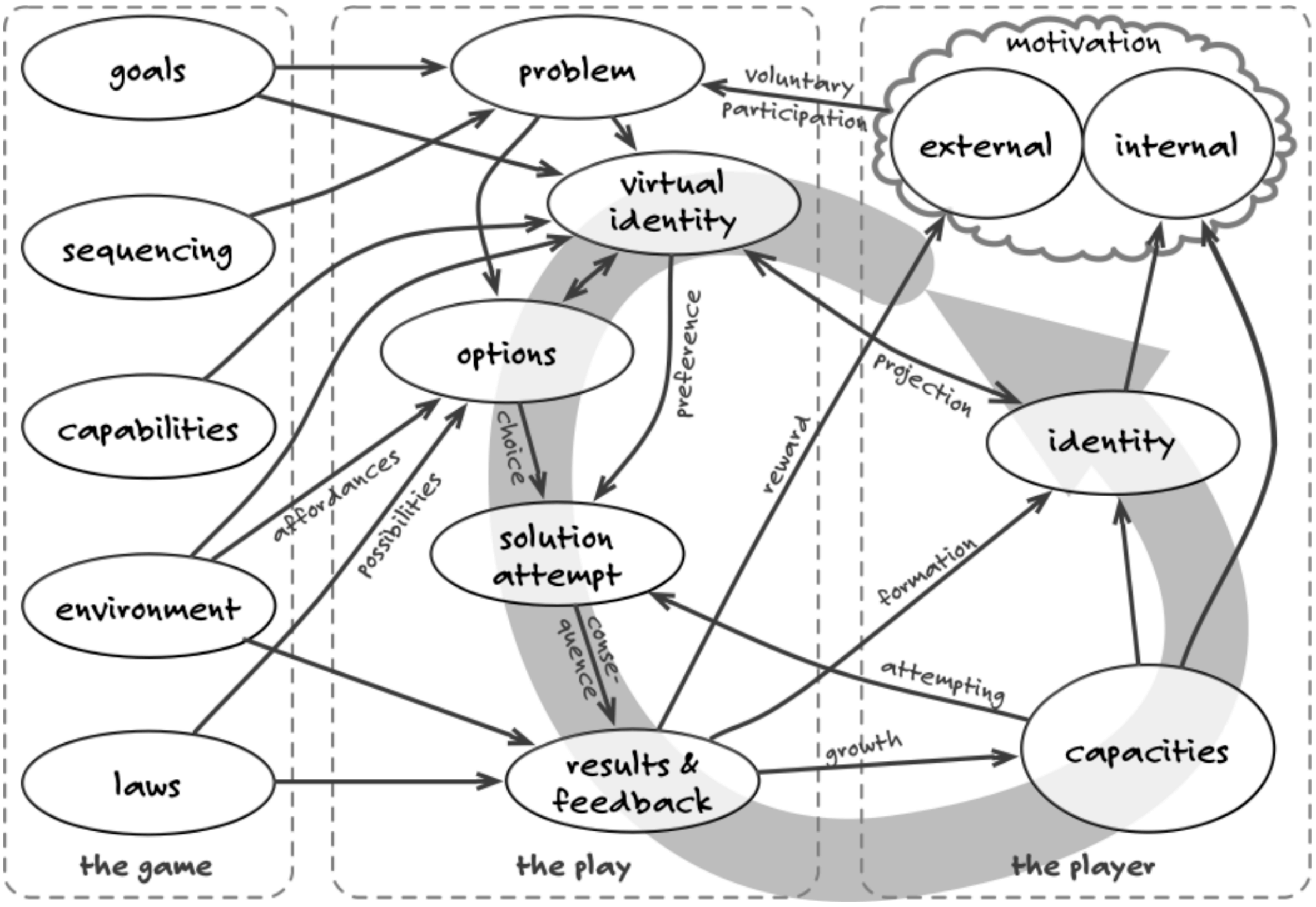

Figure 7. The *identity loop* shown on the generic framework layer.

### Future Work

This framework could be enhanced by connecting its various elements to the specific design patterns and strategies that successful video games employ, in the hope that these will suggest equivalent or similar strategies for instructional design. Additionally, a careful analysis of the ways in which it connects to or resonates with other theoretical frameworks in the learning sciences—such as identity theory or cultural-historical activity theory, for example—could both increase its utility and challenge it. The best test, however, is to see whether it can productively support design experiments using novel instructional approaches.